# Interaction description from the non-linear electromagnetic theory point of view

Alexander G. Kyriakos

*Saint-Petersburg State Institute of Technology,
St. Petersburg, Russia.*

*Present address: Athens, Greece, e-mail: agkyriak@yahoo.com*

## Abstract

In previous papers we have shown, that there is a special kind of non-linear electrodynamics (which we name Curvilinear Wave Electrodynamics - CWED), whose equations are mathematically equivalent to the equations of photons and leptons of quantum electrodynamics. The purpose of the present paper is to show that in framework of CWED the description of the interaction is also mathematically equivalent to description of the interaction in quantum physics. Another purpose of this paper is to show, that CWED allows to unify the description of interactions in physics.

## 1.0. Introduction. A modern state of the interaction description.

In previous papers [1,2] we have shown, that there is a non-linear electrodynamics - the curvilinear waves electrodynamics (CWED), whose equations are mathematically equivalent to the equations of quantum electrodynamics.

The purpose of present paper is to show, that the mathematical description of interaction in CWED are also equivalent to that both in classical and quantum mechanics.

It is known that interaction defines the most important characteristics of the matter motion and in form of the force or energy it is included in all equations of motion: equations of Newton, Schroedinger, Dirac, equations of weak and strong interactions, etc.

### 1.1. Force and energy forms of the interaction description

As it is known, interaction can be expressed as force and as energy. The force form of the description of interaction is integral, and the energy form relatively to it is differential. These forms are interconnected and can be defined one through another. In classical physics, the force is equal to a gradient of potential energy. Generally this dependence is more complex, but is also defined by the operation of differentiation. This implies the particularity of the connection of these two kinds of interaction description: the full unambiguity of transition from force to energy (and on the contrary) does not exist. For example, it is always possible to add to the energy some function (at least, a constant) so that the force value does not change.

In modern physics the most general forms of the interaction description are introduced by Lagrange and Hamilton approaches [3,4].

## 1.2. Lagrangian and Hamiltonian aproaches

### 1.2.1. Mechanical system of a rigid body (particles)

Lagrangian mechanics works generaly in n-dimensional configuration space, which includes all parameters, defining a static state of mechanical system (coordinates of particles, orientation of rigid body, etc). A point $x_\nu$ in this space draws a curve $x_\nu(t)$ in evolution ($\nu = 1,2,...,n$, where $n$ is a number of independent variables). For such curves a functional $S(x(t))$, called action, is introduced. Only those curves, on which the action reaches an extremum, correspond to real evolution (Hamilton principle).

Usually consideration is restricted to functionals of the form

$$S = \int_{t_1}^{t_2} L\, dt, \qquad (1.1)$$

with Lagrange function $L = L\left(x_\nu, \dot{x}_\nu\right)$ dependent only on generalized coordinates and velocities $x_\nu, \dot{x}_\nu$. For one material point this expression will be written down as follows

$$L = L\left(\vec{r}, \dot{\vec{r}}, t\right) = L(\vec{r}, \vec{v}, t), \qquad (1.2)$$

where $\vec{r}, \vec{v}, t$ are radius-vector, velocity and time correspondingly.

The condition of extremum for the action

$$\delta S = \delta\left(\int_{t_1}^{t_2} L\, dt\right) = 0, \qquad (1.3)$$

leads to Euler-Lagrange equations

$$\frac{d}{dt}\frac{\partial L}{\partial \dot{x}_i} - \frac{\partial L}{\partial x_i} = 0, \qquad (1.4)$$

This is (normally) a system second order differential equations, with solutions uniquely defined by initial coordinates and velocities $x_\nu(0), \dot{x}_\nu(0)$.

In Hamiltonian mechanics approach the state of the system is described by a point (*x,p*) in 2n-dimensional phase space, where *p* is momenta of particle. The dynamics is defined by a function H(*x,p*), called Hamilton function, via equations:

$$\dot{x} = \frac{\partial H}{\partial p}, \quad \dot{p} = \frac{\partial H}{\partial x}, \qquad (1.5)$$

Transition from Lagrange to Hamilton function mechanics is performed by *Legendre transformation*. It defines the momenta and Hamilton function as:

$$p\left(x, \dot{x}\right) = \frac{\partial L}{\partial \dot{x}}, \quad H(x,p) = p\dot{x} - L, \qquad (1.6)$$

The Hamilton function depends on coordinates and momenta, so one should express the velocities via momenta, inverting the definitions of momenta: $\dot{x} = \dot{x}(x,p)$, and substitute the result into Hamilton function.

### 1.2.2. Continuous systems (fields)

For fields the Lagrange function is defined by density of Lagrange function $\bar{L}$ in following way:

$$L = \int \bar{L} d\tau, \tag{1.7}$$

where $d\tau$ is an element of spatial volume. The Lagrange function density or Lagrangian depends generally on field functions and their derivatives, coordinates and time:

$$\bar{L} = \bar{L}\left(\psi_\mu, \frac{\partial \psi_\mu}{\partial x_\nu}, x_\nu\right), \tag{1.8}$$

where $\psi_\mu$ are the field functions, $\mu = 1,2,...,N$ ($N$ is a number of the functions); $\nu = 1,2,...,n$ ($n$ is a number of independent variables). In this case an action will be written down as follows:

$$S = \int_{t_1}^{t_2} L dx^1 dx^2 ... dx^n, \tag{1.9}$$

and Euler-Lagrange equations in case of the continuous system (field) become:

$$\sum_{i=1}^{n} \frac{\partial}{\partial x^\nu} \frac{\partial \bar{L}}{\partial \dot{\psi}_\mu} - \frac{\partial \bar{L}}{\partial \psi_\mu} = 0, \tag{1.10}$$

In the present time the Lagrangians are selected on base of some general requirements of symmetry (invariance), which have been produced during last century.

The approach of Hamilton in case of a continuous system performs by following way. Putting the value $\bar{H}$ named density of Hamilton function, or Hamiltonian of the system:

$$\bar{H} = \bar{H}\left(\psi_\nu, \frac{\partial \psi_\nu}{\partial x_\nu}, \pi_\nu, x_\nu\right), \tag{1.11}$$

so that

$$H = \int \bar{H} d\tau, \tag{1.12}$$

the Legendre transformation can be writen down as:

$$\pi_\nu = \frac{\partial \bar{L}}{\partial \dot{\psi}_\nu}, \quad \bar{H} = \sum_\nu \pi_\nu \dot{\psi}_\nu - \bar{L}, \tag{1.13}$$

where $\pi_\nu$ is canonical momentum density. Then the dynamics is defined by Hamiltonian via equations:

$$\dot{x} = \frac{\partial \bar{H}}{\partial \pi}, \quad \dot{\pi} = \frac{\partial \bar{H}}{\partial x}, \tag{1.14}$$

Hamilton function defines the full energy of system. When it is known, through it, by the known method, it is possible to express all other characteristics of system. This aproach is most frequently used for the description of elementary particles and fields.

## 1.3. Force form description of interaction

### 1.3.1. The conservative systems of rigid bodies

In case of conservative system of material points (i.e. of systems in which forces are gradients of potential) it has been found (postulated) that Lagrange function can be expressed as follows:

$$L = L\left(x_\nu, \dot{x}_\nu\right) = T\left(x_\nu, \dot{x}_\nu\right) - V(x_\nu), \qquad (1.15)$$

where $T\left(x_\nu, \dot{x}_\nu\right) = \sum_{\nu=1}^{n} \frac{m_\nu \vec{v}_\nu^2}{2}$ is total kinetic energy of the $n$ particles' system; $V(x_\nu) = V(x_1, x_2, ..., x_n)$ is the potential energy of this system.

This expression can be copied as:

$$L = L_{free} + L_{int}, \qquad (1.16)$$

where the first term answers to the energy of a free particle (which in this case is the kinetic energy), and the second term describes the energy of interaction of particles (in this case, potential energy).

Note, that in relativistic mechanics the correct equations of motion turn out only when instead of kinetic energy the value named kinetic potential is entered:

$$K = m_0 c^2 \left(1 - \sqrt{1 - \frac{v^2}{c^2}}\right), \qquad (1.17)$$

From Euler - Lagrange equations we obtain the equations of motion of material point (which are practically the Newton equation of motion):

$$\frac{d}{dt}\frac{\partial T}{\partial \dot{x}_i} = \frac{\partial T}{\partial x_i} - \frac{\partial V}{\partial x_i}, \qquad (1.18)$$

where $\frac{d}{dt}\frac{\partial T}{\partial \dot{x}_i} = Q_i$ are inertial forces; $\frac{\partial T}{\partial x_i} = Q_{c-K}$ are the generalized form of centrifugal and Coriolis forces; $\frac{\partial V}{\partial x_i} = Q_\nu$ are the generalized forces of interaction.

### 1.3.2. Non-conservative systems

Generally in nature the forces are not set as gradients of potential. In particular, this has no place in case of relativistic motion of bodies and in case of electrodynamics. But surprisingly, [3] here the generalized components of force can be set in such a way, that the form of the Euler - Lagrange equation is kept. Let's show it [3].

It appears, that instead of potential $V(x_\nu) = V(x_1, x_2, ..., x_n)$, which does not dependent on time, it is often possible to set the function $M = M\left(x_\nu, \dot{x}_\nu\right)$ so that the generalized force, instead of $Q_v = \dfrac{\partial V}{\partial x_i}$, can be written as:

$$Q_v' = \frac{d}{dt}\left(\frac{\partial M}{\partial \dot{x}_i}\right) - \frac{\partial M}{\partial x_i}, \tag{1.19}$$

For example in such important case as electrodynamics the Lorentz force can be expressed in the above form, if as $M$ - function we will choose the following expression:

$$M = e\left(\varphi - \frac{1}{c}\vec{v}\cdot\vec{A}\right), \tag{1.20}$$

where $\varphi$ is a scalar potential, and $\vec{A}$ is a vector potential of an electromagnetic field. Actually, substituting this expression in (1.20), we will obtain:

$$F_i = \left[\frac{d}{dt}\left(\frac{\partial}{\partial \dot{x}_i}\right) - \frac{\partial}{\partial x_i}\right]M = \left[\frac{d}{dt}\left(\frac{\partial}{\partial \dot{x}_i}\right) - \frac{\partial}{\partial x_i}\right]e\left(\varphi - \frac{1}{c}\vec{v}\cdot\vec{A}\right), \tag{1.21}$$

By differentiation of (1.21) and taking into account that $\vec{B} = \vec{\nabla}\times\vec{A}$ and $\vec{E} = -\vec{\nabla}\varphi - \dfrac{1}{c}\dfrac{\partial \vec{A}}{\partial t}$, it is easy to obtain the usual expression for Lorentz force:

$$\vec{F} = e\vec{E} + \frac{e}{c}\vec{v}\times\vec{B} = e\vec{E} + \frac{1}{c}\vec{j}\times\vec{B}, \tag{1.22}$$

Since in this case

$$L_{int} = -M\left(x_v, \dot{x}_v\right), \tag{1.23}$$

it is easy to see that the $M$ function is the energy of the electromagnetic interaction, corresponding to Lorentz force. Actually, using known relationships of 4 - current and 4 - potential:

$$j_v = ev_v = e(ic, \vec{v}), \tag{1.24}$$

$$A_v = (i\varphi, \vec{A}), \tag{1.25}$$

we obtain the known expression of the current - current interaction energy:

$$M = -j_v A_v, \tag{1.26}$$

**1.4. The modern approach to the interaction description in the quantum field theory**

Practically in the modern quantum field theory [5] there are no enough proved arguments allowing theoretically to deduce real interaction.

The rule of replacement $P_\mu$ with $P_\mu + eA_\mu$ in the presence of an electromagnetic field is known for a long time and is successfully applied to the correct description of experimental situations, when the representation of an electromagnetic field in classical potentials is meaningful. The substantiation of this choice can be made, proceeding from the gauge invariance principle. (minimal coupling).

Note anything that is important in connection with our theory [5]: identification of the phase derivative of the quantum wave function $\psi$ with the electromagnetic potential, expressed by expression

$$\frac{\partial \beta}{\partial x_\nu} = -eA_\mu, \qquad (1.27)$$

where $\beta(x)$ is a phase, leads to the existence of some observable effects, whose sense for the understanding of interaction has been realized from Aharonov and Bohm [6]. In connection with CWED it is interesting [5], that it is possible a formulation of QED without the potentials, if we recognize that non-locality is inherent to the concept of the phase, which depends on the integration way, as Mandelstam has shown [7]. Then, apparently, it is more reasonable to consider the Bohm - Aharonov experiment as the instruction on essential non-locality. (Notice that this result is full supposable in CWED framework, where $\psi$ - function belongs to an electromagnetic field).

In quantum field theory [5,8,9] it is postulated, that as well as in case of motion of rigid bodies, the Lagrangian is possible to be presented as the sum of Lagrangian of free particles, and Lagrangian of their interactions:

$$\overline{L} = \overline{L}_{free} + \overline{L}_{int}, \qquad (1.28)$$

Here Lagrangian of free particles represents the sum of Lagrangians for each free particle separately. For example, in QED we have

$$\overline{L}_{free} = \overline{L}_e + \overline{L}_\gamma, \qquad (1.29)$$

where $\overline{L}_e, \overline{L}_\gamma$ is Lagrangians of free electron and photon accordingly. The Lagrangian of the interaction is either postulated, or obtained due to gauge invariance in the form of current-current interaction.

As it is known in the theory of Standard Model (SM) the Lagrangians of free particles and their interactions are a generalization of Lagrangian of QED [8, 9].

The target of the following part of the paper is to show that CWED allows us to obtain the mathematical description of interaction, which is adequate to the modern description of interactions.

Since in framework of CWED the fields of particles and the particles themselves are electromagnetic and only electromagnetic, further we will consider the fields and particles of CWED generally as electromagnetic fields and will name them "the CWED particles".

## 2.0. Lagrangian and Hamiltonian of the linear electromagnetic field

As it is known [4,10,11] the Lagrangian and Hamiltonian, which completely defines the system of interacting fields (photons) in the classical electrodynamics, looks like:

$$\overline{L} = \frac{1}{8\pi}\left(\vec{E}^2 - \vec{H}^2\right), \qquad (2.1)$$

$$\overline{H} = \frac{1}{8\pi}\left(\vec{E}^2 + \vec{H}^2\right), \qquad (2.2)$$

where $\vec{E}$ and $\vec{H}$ are electric and magnetic fields.

Let's show, that this Lagrangian allows us also to describe the interactions as it has place in modern field theory. These results are convenient to unify in the following theorem:

*Lagrangian (Hamiltonian) of the system of interaction fields (particles), due to the principle of fields' superposition, is automatically divided on two parts:*
*1. Lagrangian of the free particles, which is equal to the sum of Lagrangians, each of which describes the Lagrangian of one free particle and does not depend on the presence of other particles. Lagrangian of one free particle is determined by squares of own fields of particle,*
*2. Lagrangian of the interaction of particle, which is equal to the sum of Lagrangians, each of which describes interaction of one pair particles of system and does not depend on presence of other particles. Lagrangian of the interaction of any two particles is described by cross products of fields of these two particles.* Let's try to prove this theorem.

Let the system consists of two parts (particles) 1 and 2, which have both the electric and magnetic fields: $\vec{E}_1$, $\vec{H}_1$ and $\vec{E}_2$, $\vec{H}_2$. According to a principle of superposition a total field of system of particles is equal to the sum of the fields created by each particle separately: $\vec{E} = \vec{E}_1 + \vec{E}_2$, $\vec{H} = \vec{H}_1 + \vec{H}_2$.

Thus, for Lagrangian and Hamiltonian of two interacting particles we obtain:

$$\overline{L} = \frac{1}{8\pi}\left(\vec{E}^2 - \vec{H}^2\right) = \frac{1}{8\pi}\left(\vec{E}_1^2 - \vec{H}_1^2\right) + \frac{1}{8\pi}\left(\vec{E}_2^2 - \vec{H}_2^2\right) + \frac{1}{8\pi}\left(\vec{E}_1\vec{E}_2 - \vec{H}_1\vec{H}_2\right), \quad (2.3)$$

$$\overline{H} = \frac{1}{8\pi}\left(\vec{E}^2 + \vec{H}^2\right) = \frac{1}{8\pi}\left(\vec{E}_1^2 + \vec{H}_1^2\right) + \frac{1}{8\pi}\left(\vec{E}_2^2 + \vec{H}_2^2\right) + \frac{1}{8\pi}\left(\vec{E}_1\vec{E}_2 + \vec{H}_1\vec{H}_2\right), \quad (2.4)$$

As we see, Lagrangian of systems of two interacting particles is actually broken into two parts, one of which is defined only by own fields of particles, and the other is defined only by fields of pair of different particles.

In case that the particle is a photon, it is easy to see that terms of the first part correspond to Lagrangians of free particles [11]. In case the particle is electron-like particle of CWED, the theorem is proved below in the section 4.0, devoted to Lagrangian of boson-like and electron-like particles.

Thus, we need to prove now that these terms define the interaction of the charge particles.

## 2.1. The interaction description of two charge particles

Taking into account the known results [4,10,11,12] we will prove the above theorem for the case of two charged particles.

But it is not difficult to see that due to the fact that the general Lagrangian of system of particles is defined by a square of the sum of fields, in case of any number of charge particles the *cross terms will be defined by fields that belong only to two different particles*.

## 2.1.1. The description of interaction in case of static particles.

Let us consider first a case when only an electrostatic field is present.

Let we have two charges $q_1$ and $q_2$, situated on distance $r_a$ from each other. The values of a field in any point of space P, which are situated on distances with radius-vectors $\vec{r}_1$ and $\vec{r}_2$ from charges, are defined by expressions:

$$\vec{E}_1 = \frac{q_1}{r_1^2}\vec{r}_1^{\,0}, \quad \vec{E}_2 = \frac{q_2}{r_2^2}\vec{r}_2^{\,0}, \quad (2.5)$$

where $\vec{r}_1^{\,0}$ and $\vec{r}_2^{\,0}$ are the unit vectors of corresponding radius-vectors, $r_1$ and $r_2$ are their absolute values. The energy density of an electric field in point P is equal to a square of a vector of an electric field in this point:

$$\frac{1}{8\pi}\vec{E}^2 = \frac{1}{8\pi}\left(\vec{E}_1+\vec{E}_2\right)^2 = \frac{1}{8\pi}\left[\vec{E}_1^2 + 2\vec{E}_1\vec{E}_2 + \vec{E}_2^2\right] = \frac{q_1^{\,2}}{8\pi\, r_1^{\,4}} + \frac{q_2^{\,2}}{8\pi\, r_2^{\,4}} + 2\frac{q_1 q_2}{8\pi\, r_1^{\,2} r_2^{\,2}}\cos\theta, \quad (2.6)$$

where $\theta$ is a angle between vectors $\vec{r}_1$ and $\vec{r}_2$. Thus the Lagrangian of a total field can be written down as:

$$\overline{L} = \overline{L}_{o1} + \overline{L}_{o2} + \overline{L}_{12}, \quad (2.7)$$

where $\overline{L}_{o1} = \frac{1}{8\pi}\vec{E}_1^2 = \frac{q_1^{\,2}}{8\pi\, r_1^{\,4}}$, $\overline{L}_{o2} = \frac{1}{8\pi}\vec{E}_2^2 = \frac{q_2^{\,2}}{8\pi\, r_2^{\,4}}$, $\overline{L}_{12} = \frac{1}{4\pi}\vec{E}_1 \cdot \vec{E}_2 = 2\frac{q_1 q_2}{8\pi\, r_1^{\,2} r_2^{\,2}}\cos\theta$.

Here the first and second terms, obviously, represent the Lagrangian of free particles (fields). To find out the meaning of third term, containing cross product of charges $q_1 q_2$, we will calculate the Lagrange function, corresponding to this term, using (1.7). As $\vec{E}_2 = -grad\varphi_2 = -\vec{\nabla}\varphi_2$ where $\varphi_2 = \frac{q_2}{r_2}$ is static potential for the second charge, we will obtain:

$$L = -\frac{1}{4\pi}\int \vec{\nabla}\varphi_2 \cdot \vec{E}_1 d\tau, \quad (2.8)$$

Integrating by parts, we obtain:

$$L = -\frac{1}{4\pi}\int \vec{\nabla}\varphi_2 \cdot \vec{E}_1 d\tau = -\frac{1}{4\pi}\varphi_2(E_x + E_y + E_z)\Big|_{-\infty}^{\infty} + \frac{1}{4\pi}\int \varphi_2(\vec{\nabla}\vec{E})d\tau, \quad (2.9)$$

Here the first term is equal to zero, and in the second term, according to Maxwell we have:

$$\vec{\nabla}\vec{E} = 4\pi\rho_e, \quad (2.10)$$

where $\rho_e$ is the density of an electric charge $q_1$. Then, accepting, that $r_o \ll r_a$, we obtain:

$$\overline{L}_{12} = \rho_e \varphi, \quad (2.11)$$

This means, that *the Lagrangian, adequate to the third term, is Lagrangian of interaction of two charges and has the form of a current - current interaction for the case of a static field.*

### 2.1.2. The description of interaction in case of moving particles

Now we will consider the Lagrangian of two interacting charges, which are in motion. Here alongside with an electric field the magnetic field will also appear. Thus, we should analyze a general view of the Lagrangian in case of any motion of electric charges

First of all, a question arises, of whether the electric field varies in case that charges move. This question can be formulated in more general sense: will the Gauss theorem be fair in case the charges move? The experiment answers positively [13]. Hence, the above-stated analysis, concerning an static electric field, will be fair as well as in case of moving charges.

Thus, further it is enough to analyze only the term of the general Lagrangian, which contains a magnetic field.

In point P, the magnetic fields from each particle have the form:

$$\vec{H}_1 = \frac{q_1}{r_1^{\,2}}\left[\vec{v}\times\vec{r}_1^{\,0}\right], \qquad \vec{H}_2 = \frac{q_2}{r_2^{\,2}}\left[\vec{v}\times\vec{r}_2^{\,0}\right], \quad (2.12)$$

(where $\vec{v}$ is the particle velocity) and the energy density will be

$$\frac{1}{8\pi}\vec{H}^2 = \frac{1}{8\pi}(\vec{H}_1 + \vec{H}_2)^2 = \frac{1}{8\pi}[\vec{H}_1^2 + 2\vec{H}_1\vec{H}_2 + \vec{H}_2^2] =$$
$$= \frac{q_1^2}{8\pi\, r_1^4}[\vec{v}\times\vec{r}_1^{\,0}]^2 + \frac{q_2^2}{8\pi\, r_2^4}[\vec{v}\times\vec{r}_2^{\,0}]^2 + 2\frac{q_1 q_2}{8\pi\, r_1^2 r_2^2}[\vec{v}\times\vec{r}_1^{\,0}]\cdot[\vec{v}\times\vec{r}_2^{\,0}]$$
(2.13)

Using (2.13) for Lagrangian we have

$$\overline{L} = \overline{L}_{o1} + \overline{L}_{o2} + \overline{L}_{12},$$
(2.14)

where $\overline{L}_{o1} = \frac{1}{8\pi}\vec{H}_1^2$, $\overline{L}_{o2} = \frac{1}{8\pi}\vec{H}_2^2$ are Lagrangian of free particles, and the term $\overline{L}_{12} = \frac{1}{4\pi}\vec{H}_1\cdot\vec{H}_2$ is according to our supposition the Lagrangian of interaction.

We will show it by calculating of the Lagrangian:

$$L_{12} = \int \overline{L}_{12} d\tau,$$
(2.15)

As $\vec{H}_2 = \vec{\nabla}\times\vec{A}_2$, where $\vec{A}_2 = \frac{1}{c}\frac{i_2}{r_2}\vec{r}_2^{\,o} = \frac{1}{c}\int\frac{\vec{j}_2}{r_2}d\tau_2$ is the vector potential of current of the second charge (here $\vec{j}$ is the current density), we will obtain:

$$L_{12} = \frac{1}{4\pi}\int \vec{\nabla}\times\vec{A}_2 \cdot \vec{H}_1 d\tau,$$
(2.16)

Integrating by parts in scalar form, we obtain:

$$L_{12} = \frac{1}{4\pi}\sum_{l,n}(\pm)\vec{A}_{2l}\vec{H}_m\Big|_{-\infty}^{+\infty} + \frac{1}{4\pi}\int \vec{A}_2 (\vec{\nabla}\times\vec{H})d\tau,$$
(2.17)

where $l = (x, y, z)$, $m = (x, y, z)$, $l \ne m$ and under the sum the signs are alternated. Here the first term is equal to zero, and in the second term according to Maxwell we have:

$$\vec{\nabla}\times\vec{H} = \frac{4\pi}{c}\vec{j},$$
(2.19)

Then we obtain:

$$\overline{L}_{12} = \overline{L}_{int} = \frac{1}{c}\vec{j}\cdot\vec{A},$$
(2.20)

This means that the *Lagrangian, relatively to magnetic fields of two moving charges, has the form of current - current interaction.*

So, generally we obtain that the interaction Lagrangian of two moving charge particles is defined by the commutator of the electric and magnetic fields of two particles, and can be written down in the form of a current - current interaction:

$$\overline{L}_{int} = -j_\nu A_\nu,$$
(2.21)

Obviously, in this case the general Hamiltonian of interactions will be written down as follows:

$$H = -\frac{1}{2}\int(\rho\varphi + \vec{j}\cdot\vec{A})\,d\tau,$$
(2.22)

## 2.2. Interaction description of systems of charge particles

In case that the system consists of a number of charged particles of both signs and different sizes, we receive the object, possessing various new electromagnetic properties.

As it is known, the system of moving charges possesses in general the electromagnetic moments. In some cases the total charge or current of a system (which in this case are also named zero moments of a system) can be equal to zero (i.e. it is a neutral system) while other moments are not equal to zero. This means, that these systems are capable of interacting due to other moments. But the interaction energy of such systems is much lower than the energy of interaction of the charged systems, but is not equal to zero. For example, the atoms, being the neutral objects, nevertheless are capable of interacting among themselves by various forces, which frequently are named Van der Waals forces (these forces are weaker than Coulomb force and depend in inverse proportion on five or six power of distance). In QM these forces depend additionally on spin orientation and other quantum parameters.

We have shown that the neutral particle in framework of CWED [2] is also an electromagnetic field. Thus, in case of neutral particles the interaction must also described by the formula of a current - current interaction.

For the charge system description potentials are usually used. As it is known, the use of potentials facilitates the mathematical analysis of problems of electrodynamics. Since in the CWED instead of potentials, the strength of electromagnetic field is considered as wave function, we remember that in electrodynamics there is an opportunity to write the interaction through field strengths.

As it is known [10,11], the electromagnetic fields of a moving charge can be described as following:

$$\vec{E}(\vec{r},t) = \frac{e(1-\vec{v}^2/c^2)(\vec{n}-\vec{v}/c)}{R^2(1-\vec{n}\cdot\vec{v}/c)^3}\bigg|_{t'} + \frac{e\vec{n}\times[(\vec{n}-\vec{v}/c)\times\dot{\vec{v}}]}{R^2(1-\vec{n}\cdot\vec{v}/c)^3}\bigg|_{t'}, \qquad (2.23)$$

$$\vec{H}(\vec{r},t) = \frac{e(1-\vec{v}^2/c^2)(\vec{v}\times\vec{n})}{R^2(1-\vec{n}\cdot\vec{v}/c)^3}\bigg|_{t'} + \frac{e\{c\dot{\vec{v}}\times\vec{n}+\vec{n}\times[(\vec{v}\times\dot{\vec{v}})\times\vec{n}]\}}{R^2(1-\vec{n}\cdot\vec{v}/c)^3}\bigg|_{t'}, \qquad (2.24)$$

Each of these expressions consists of two components. The first component forms quasi-stationary fields, which change in space as $R^{-2}$ and does not contain the acceleration of a charge. The second component describes a wave field of radiation: it is proportional to acceleration $\dot{\vec{v}}$ and decreases as $R^{-1}$. It is easy to show, that the energy flux of quasi-stationary fields decreases as $R^{-2}$. Hence, the quasi-stationary field remains all time connected with a particle and does not create energy flux on infinity.

Thus, on large distances from a particle in expressions (2.23)-(2.24) only the second terms (named wave field) remains. This means that the electromagnetic perturbations can propagate from charge particle to the infinity. Due to these fields, i.e. electromagnetic waves, the particle systems interact with each other on a long distance. Obviously in linear case this interaction is represented by the interference of electromagnetic waves.

The description of charged particles' system by means of potentials is more often, although it doesn't have any advantages. In particular, [10,11] for the scalar potential in large distances from the system of charges, we have the expansion:

$$\varphi(r) = \frac{q}{r} + \frac{\vec{p} \cdot \vec{r}}{r^3} + \frac{Q_{\alpha\beta} x_\alpha x_\beta}{2r^5} + \ldots, \qquad (2.25)$$

where in case of continuous distribution of charges we have: $q = \int \rho(r')dV'$ is the full charge of the system, $\vec{p} = \int r' \rho(r')dV'$ is the dipole moment of the system, $Q_{\alpha\beta} = \int \rho(r')(3x'_\alpha x'_\beta - r' \delta_{\alpha\beta})dV'$ is the tensor of the quadrupole moment of the system of charges, etc. (in case of a discrete system of point charges we have sums instead of integrals).

In general case of arbitrarily moving charges we obtain the so-called Lienar - Wiechert potential [10,11]s:

$$\vec{A}(\vec{r},t) = \frac{e\vec{v}}{cR(1-\vec{n}\cdot\vec{v}/c)}\bigg|_{t'}, \qquad (2.26)$$

$$\varphi(\vec{r},t) = \frac{e}{cR(1-\vec{n}\cdot\vec{v}/c)}\bigg|_{t'}, \qquad (2.27)$$

where $\vec{s}(t')$ are the coordinates of the particle, $\vec{r}$ are the coordinates of the observation point, $\vec{v}(t) = \dot{\vec{s}}(t)$ is its velocity, $e$ is the charge, $\vec{R}(t') = \vec{r} - \vec{s}(t')$ and $t'$ is the retarded moment of time, which is defined by the relation:

$$c(t-t') = |\vec{r} - \vec{s}(t')|, \qquad (2.28)$$

so that difference $t - t' = |\vec{r} - \vec{s}(t')|/c$ represents the time of distribution of the electromagnetic perturbation from the particle up to an observation point of the field.

Thus, in case of the neutral particles the interaction Lagrangian (Hamiltonian) doesn't equal to zero, but can be presented as series, some terms of which will be equal to zero.

## 3.0. Consequences of the theorem

### 3.1. About masses of the interaction particles Transforming Energy into Mass

We have shown above that the cross product of fields in Lagrangian accords to a current - current interaction form. From this, the next important consequences follow.

**1**. *The energy of two or more interacting objects is bigger than the energy of free objects, and the difference corresponds to the term of cross product of fields.*

An important question arises: how the interaction energy of two objects is divided between them? Let's consider one concrete case. For simplicity we will only talk about an electric field. The full density of energy of two interacting particles looks like:

$$u = \frac{1}{8\pi}(\vec{E}_1 + \vec{E}_2)^2 = u_{1o} + u_{int} + u_{2o}, \qquad (3.1)$$

where $u_{1o} = \frac{1}{8\pi}\vec{E}_1^2$ and $u_{2o} = \frac{1}{8\pi}\vec{E}_2^2$ are the densities of energy of the first and second particles in a free state accordingly, and $u_{int} = \frac{1}{8\pi}[\vec{E}_1\vec{E}_2 + \vec{E}_2\vec{E}_1]$ is the density of interaction energy of these particles. As both components in the above formula are equal, we can accept that:

**2.** *The interaction energy is divided fifty-fifty between two interacting particles.*

From above it follows that the interaction energy density of each particle is equal to each other:

$$u_{int1} = \frac{1}{8\pi} \vec{E}_1 \vec{E}_2 = u_{int2}, \qquad (3.2)$$

and the full density of energy of each interacting particles is equal to:

$u_1 = \frac{1}{8\pi}\left(\vec{E}_1^2 + \vec{E}_1 \vec{E}_2\right)$ for the first particle, and $u_2 = \frac{1}{8\pi}\left(\vec{E}_2^2 + \vec{E}_1 \vec{E}_2\right)$ for the second particle.

Since the energy is defined by integral from energy density $\varepsilon = \frac{1}{c^2}\int_0^\infty u d\tau$, the same conclusion refers also to the energy.

**3**. From above, in conformity with the known expression $m = \frac{\varepsilon}{c^2}$, it follows that *the mass of each interacting particle increases in comparison with the mass of the free particle on half value of the term of the field cross product.*

In other words, the mass of interaction of two particles divides fifty-fifty between them so that $m_{int1} = m_{int2}$, and for the masses of interacting particles we have

$$m_1 = m_{01} + m_{int1}, \quad m_2 = m_{02} + m_{int2}, \qquad (3.3)$$

### 3.2. The Newton force form of interaction description

Let's show, that in framework of CWED the consequences of the above-stated theorem are the equations of classical mechanics [3].

#### 3.2.1. The second law of Newton

According to the second law of Newton we have

$$\frac{d\vec{p}}{dt} = \sum \vec{F}_i, \qquad (3.4)$$

where $\vec{p}$ is a momenta of the particle motion, $\vec{F}_i$ are all forces acted on the particle.

For particles in electromagnetic field, it is easy to show that the equation (3.4) follows from relativistic Lagrangian of the charge particle in the electromagnetic fields (see the textbooks on electrodynamics [4,11]).

If the question is only about the electromagnetic field without particles, in this case, as it is known [11,14], "The momentum theorem" exists, which shows the validity of the law of Newton concerning the electromagnetic field as a material carrier (note that in framework of CWED "The momentum theorem" can easy to prove [15] by other way).

We should also remind that according to Ehrenfest theorem [16] it is possible from Dirac equation to obtain the equation of Newton (3.4). As the Dirac equation coincides precisely with the electron-like equation of CWED [1], it is possible to assert that Ehrenfest theorem proves that the equation of Newton also follows from CWED [15].

*3.2.2. The third law of Newton.*

Recall the Newton's Third Law of motion: "For every action there is an equal and opposite reaction". It is easy to make sure, that the third law of Newton also follows from the above-stated conclusions. Actually, using the expression (1.20), we can write down:

$$\varepsilon_{int1} = \varepsilon_{int2} = \frac{1}{2} M, \tag{3.5}$$

Using the Euler-Lagrange approach (see (1.21)) it is easy to show from (3.5) that action and counteraction forces are equal. In addition, depending on which particle we will choose as initial, the direction of force will be opposite in relation to the other choice. Thus we obtain Newton's Third Law of motion: $\vec{F}_1 = -\vec{F}_2$

## 4.0. Lagrangian of interaction of CWED particles

Here we will consider the electrodynamics form of the Lagrangians of the **CWED** vector boson and the **CWED** lepton equations.

### 4.1. Lagrangian of the vector boson-like particles

Let us consider the plane electromagnetic wave moving on $y$- axis [1,2]. In the general case it has two polarizations and contains the following field vectors:

$$E_x, E_z, H_x, H_z, \tag{4.1}$$

(As it is known, the relation $E_y = H_y = 0$ takes place for all transformations, so that there are always only four components (4.1) as in the Dirac $\psi$-function).

Let's enter the electromagnetic wave fields as the Dirac bispinor matrix:

$$\psi = \begin{pmatrix} E_x \\ E_z \\ iH_x \\ iH_z \end{pmatrix}, \quad \psi^+ = \begin{pmatrix} E_x & E_z & -iH_x & -iH_z \end{pmatrix}, \tag{4.2}$$

In this case $\psi = \psi(y)$ (for all other directions of the electromagnetic waves the matrices choice have been considered in the previous paper [1]).

We can write the wave equation in the following form:

$$\left(\hat{\varepsilon}^2 - c^2 \hat{p}^2\right)\psi' = 0, \tag{4.3}$$

where $\hat{\varepsilon} = i\hbar \frac{\partial}{\partial t}$, $\hat{\vec{p}} = -i\hbar \vec{\nabla}$ are the operators of energy and momentum, respectively.

Using (4.2), we can prove that (4.3) is also the equation of the electromagnetic wave, moving along the $y$- axis. Taking into account that

$$\left(\hat{\alpha}_o \hat{\varepsilon}\right)^2 = \hat{\varepsilon}^2, \quad \left(\hat{\vec{\alpha}} \cdot \hat{\vec{p}}\right)^2 = \hat{\vec{p}}^2, \tag{4.4}$$

the equation (4.3) can also be written in the following form:

$$(\hat{\alpha}_o \hat{\varepsilon})^2 - c^2 (\hat{\vec{\alpha}} \cdot \hat{\vec{p}})^2 \psi' = 0, \tag{4.5}$$

where $\hat{\alpha}_0 = \begin{pmatrix} \hat{\sigma}_0 & 0 \\ 0 & \hat{\sigma}_0 \end{pmatrix}$; $\hat{\vec{\alpha}} = \begin{pmatrix} 0 & \hat{\vec{\sigma}} \\ \hat{\vec{\sigma}} & 0 \end{pmatrix}$; $\hat{\beta} \equiv \hat{\alpha}_4 = \begin{pmatrix} \hat{\sigma}_0 & 0 \\ 0 & -\hat{\sigma}_0 \end{pmatrix}$ are Dirac's matrices, and $\hat{\vec{\sigma}}$ are Pauli matrices $\hat{\sigma}_x = \begin{pmatrix} 0 & 1 \\ 1 & 0 \end{pmatrix}$, $\hat{\sigma}_y = \begin{pmatrix} 0 & -i \\ i & 0 \end{pmatrix}$, $\hat{\sigma}_z = \begin{pmatrix} 1 & 0 \\ 0 & -1 \end{pmatrix}$, $\hat{\sigma}_0 = \begin{pmatrix} 1 & 0 \\ 0 & 1 \end{pmatrix}$.

Factorizing (4.4) we get:

$$(\hat{\alpha}_o \hat{\varepsilon} - c\hat{\vec{\alpha}} \cdot \hat{\vec{p}})(\hat{\alpha}_o \hat{\varepsilon} + c\hat{\vec{\alpha}} \cdot \hat{\vec{p}})\psi' = 0, \tag{4.6}$$

As it is follows from previous paper [1], due to the curvilinear motion of the electromagnetic wave, some additional terms $K$, corresponding to the tangent components of the displacement current, will appear in the equation (4.6), so that from (4.6) we have:

$$(\hat{\alpha}_o \hat{\varepsilon} - c\hat{\vec{\alpha}} \cdot \hat{\vec{p}} - K)(\hat{\alpha}_o \hat{\varepsilon} + c\hat{\vec{\alpha}} \cdot \hat{\vec{p}} + K)\psi = 0, \tag{4.7}$$

where $K = \hat{\beta} mc^2$ and $mc^2$ is half of electromagnetic wave (photon) energy. Thus, in the case of the curvilinear motion of the electromagnetic field (photon) instead of the equation (2.6) we obtain the Klein-Gordon-like equation with mass:

$$(\hat{\varepsilon}^2 - c^2 \hat{p}^2 - m^2 c^4) \psi = 0, \tag{4.8}$$

Note that $\psi$-function, which appears after electromagnetic wave twirling, is not identical to the $\psi'$-function before twirling: the $\psi'$-function is the classical electromagnetic wave field and the $\psi$-function is the "electromagnetic spinor" field (see in detail [1,2])

As it is known the Klein-Gordon equation is considered as the scalar field equation. But obviously the Klein-Gordon equation (4.8), whose wave function is 4-matrix with electromagnetic field components, cannot have the sense of the scalar field equation. Really, let us analyze the objects, which this equation describes.

From the Maxwell equations follows, that each of the components $E_x, E_y, E_z, H_x, H_y, H_z$ of vectors of an electromagnetic field $\vec{E}, \vec{H}$ submits to same form of the scalar wave equation. In this case if it is required to know the changing of one of the vectors components $\vec{E}, \vec{H}$ only, we can consider the vector field as scalar. But in case of CWED, when a tangential current appears, we cannot proceed to the scalar theory, since the components of a vector $\vec{E}$, as it follows from the condition $\vec{\nabla} \cdot \vec{E} = \frac{4\pi}{c} \vec{c}^0 \cdot \vec{j}$, are not independent functions (here $\vec{c}^o$ is the unit vector of wave velocity).

Therefore, although the scalar wave equations (4.8) represent Klein - Gordon equations, the equation (4.8) concerning function (4.2) after curvilinear transformation represents the equation of the vector particle with mass, which we can name "heavy" photon. Whereas the Klein - Gordon equation describes a massive particle with spin zero (spinless boson), the equation (4.8) describes the twirled photon, i.e. a massive particle with unit spin (i.e. the particle similar to the intermediate boson).

The Lagrangian of the equation (4.8) can be written in the form:

$$L = \psi^+ (\hat{\alpha}_o \hat{\varepsilon} - c\hat{\vec{\alpha}} \cdot \hat{\vec{p}} + \hat{\beta} mc^2)(\hat{\alpha}_o \hat{\varepsilon} + c\hat{\vec{\alpha}} \cdot \hat{\vec{p}} - \hat{\beta} mc^2) \psi, \tag{4.9}$$

or

$$L = \partial_\mu \psi^+ \partial^\mu \psi - m^2 c^4 \psi^+ \psi, \tag{4.10}$$

### *4.1.2. Lagrangian of the lepton-like particles*

The Lagrngians of the fermions can write in the following form [16]:

$$L = \psi^+ \left( \hat{\alpha}_o \hat{\varepsilon} \mp c\hat{\vec{\alpha}} \cdot \hat{\vec{p}} \pm \hat{\beta} mc^2 \right) \psi, \tag{4.11}$$

The Lagrangian (4.11) can be represents as
$$L = L_0 + L', \tag{4.12}$$

where

$$L_0 = \psi^+ \left( \hat{\alpha}_o \hat{\varepsilon} \mp c\hat{\vec{\alpha}} \cdot \hat{\vec{p}} \right) \psi, \tag{4.13}$$

and

$$L' = \psi^+ \left( \pm \hat{\beta} mc^2 \right) \psi, \tag{4.14}$$

Note that due to Dirac equation we have $L = \psi^+ \left( \hat{\alpha}_o \hat{\varepsilon} \mp c\hat{\vec{\alpha}} \cdot \hat{\vec{p}} \pm \hat{\beta} mc^2 \right) \psi = 0$ and can write

$$\psi^+ \left( \pm \hat{\beta} mc^2 \right) \psi = \psi^+ \left( \hat{\alpha}_o \hat{\varepsilon} \mp c\hat{\vec{\alpha}} \cdot \hat{\vec{p}} \right) \psi, \tag{4.15}$$

Let us analyze the electromagnetic sense of each part of (4.15) using only minus sign.

Let consider the Euler-Lagrange equation, which corresponds to (4.11). Taking into account (4.2) we obtain the following equation:

$$\begin{cases} rot\ \vec{E} + \dfrac{1}{c} \dfrac{\partial \vec{H}}{\partial t} = i\dfrac{\omega}{c} \vec{H}, \\ rot\ \vec{H} - \dfrac{1}{c} \dfrac{\partial \vec{E}}{\partial t} = i\dfrac{\omega}{c} \vec{E}, \end{cases} \tag{4.16}$$

where $\omega = \dfrac{mc^2}{\hbar}$. The equation (4.16) is the Maxwell equations with imaginary currents, and these currents have [1,2] the form of the tangent component of the displacement current of the twirled electromagnetic wave:

$$\vec{j}_\tau^e = i\dfrac{\omega}{4\pi} \vec{E}, \quad \vec{j}_\tau^m = i\dfrac{\omega}{4\pi} \vec{H}, \tag{4.17}$$

It is not difficult to see that the Lagrangian (4.13) defines the Maxwell equation without current, i.e. this part is the Lagrangian of a free electromagnetic wave (photon).

For other term, in the electromagnetic form we have
$$\psi^+ \hat{\beta} \psi = \vec{E}^2 - \vec{H}^2, \tag{4.18}$$

Using (4.17) the expression (4.18) can be presented in other form: (5.74)

$$L' = i\dfrac{\omega_e}{8\pi} \left( \vec{E}^2 - \vec{H}^2 \right) = \dfrac{1}{2} \left( \vec{j}_\tau^e \vec{E} - \vec{j}_\tau^m \vec{H} \right), \tag{4.19}$$

Thus, this part of the Lagrangian of lepton is the self-action Lagrangian and is responsible for interaction of its own lepton current with its own lepton field.

It is easy to generalize this expression in the case of the lepton interaction with an external field of other particle. According to the superposition principle of the fields, we have:

$$\vec{E} = \vec{E}_{in} + \vec{E}_{ex}, \quad \vec{H} = \vec{H}_{in} + \vec{H}_{ex}, \tag{4.20}$$

Where, with the index "in" we have designated the characteristics of the investigated particle, which we described above without an index. And the index "ex" is used for the particle, which is "external" in relation to the "in"- particle.

Substituting (4.20) in (4.19), we obtain:

$$L' = i\frac{\omega}{8\pi}\left[(\vec{E}_{in}^2 - \vec{H}_{in}^2) + 2(\vec{E}_{in}\vec{E}_{ex} - \vec{H}_{in}\vec{H}_{ex}) + (\vec{E}_{ex}^2 - \vec{H}_{ex}^2)\right], \tag{4.21}$$

The terms of expression (4.21) in case of interaction of particles are possibly presented as follows:

$$i\frac{\omega}{8\pi}(\vec{E}_{in}^2 - \vec{H}_{in}^2) = \frac{1}{2}(\vec{j}_{in}{}^e \vec{E}_{in} - \vec{j}_{in}{}^m \vec{H}_{in})$$

$$i\frac{\omega}{4\pi}(\vec{E}_{in}\vec{E}_{ex} - \vec{H}_{in}\vec{H}_{ex}) = (\vec{j}_{in}{}^e \vec{E}_{ex} - \vec{j}_{in}{}^m \vec{H}_{ex}) = (\vec{j}_{ex}{}^e \vec{E}_{in} - \vec{j}_{ex}{}^m \vec{H}_{in}), \tag{4.22}$$

$$i\frac{\omega}{8\pi}(\vec{E}_{ex}^2 - \vec{H}_{ex}^2) = \frac{1}{2}(\vec{j}_{ex}^e \vec{E}_{ex} - \vec{j}_{ex}^m \vec{H}_{ex})$$

The first of the equations (4.22) corresponds to the self-action of the "in"- particle, i.e. to interaction of own current of this particle with its own field.

It is easy to see, that the second expression from (4.22) describes the interaction of the "in"- particle with the "ex"- particle. As we have shown above, the cross product of vectors of a field is responsible for the interaction of particles and leads to the current-current interaction. Using the expression $(\vec{j}_{ex}^e \vec{E}_{in} - \vec{j}_{ex}^m \vec{H}_{in})$, it is not difficult to obtain the Dirac equation of lepton in an external field (see below).

Obviously, the last expression from (4.22) describes the self-action of "ex"- particles and does not play any role in the description of "in"- particle.

## 5.0 About mass of interacting particles

Let us consider the Dirac equation with an external field:

$$\left[(\hat{\alpha}_o\hat{\varepsilon} - c\hat{\vec{\alpha}}\cdot\hat{\vec{p}}) + (\hat{\alpha}_o\varepsilon_{ex} - c\hat{\vec{\alpha}}\cdot\vec{p}_{ex}) + \hat{\beta}\ mc^2\right]\psi = 0, \tag{5.1}$$

As the internal field can be expressed through the electron mass:

$$\hat{\beta}\ mc^2 = \hat{\alpha}_o\varepsilon_{in} - c\hat{\vec{\alpha}}\cdot\vec{p}_{in}, \tag{5.2}$$

we can write (5.1) as:

$$\left[(\hat{\alpha}_o\hat{\varepsilon} - c\hat{\vec{\alpha}}\cdot\hat{\vec{p}}) + (\hat{\alpha}_o\varepsilon_{ex} - c\hat{\vec{\alpha}}\cdot\vec{p}_{ex}) + (\hat{\alpha}_o\varepsilon_{in} - c\hat{\vec{\alpha}}\cdot\vec{p}_{in})\right]\psi = 0 \tag{5.3}$$

According to (5.2) we can say, that some additional mass corresponds to an external field:

$$\hat{\alpha}_o\varepsilon_{ex} - c\hat{\vec{\alpha}}\cdot\vec{p}_{ex} = \hat{\beta}\ m_{ad}c^2 \tag{5.4}$$

Using (5.4), we will obtain:

$$\left[(\hat{\alpha}_o\hat{\varepsilon} - c\hat{\vec{\alpha}}\cdot\hat{\vec{p}}) + \hat{\beta}\ (m + m_{ad})c^2\right]\psi = 0 \tag{5.5}$$

or

$$(\hat{\alpha}_o\hat{\varepsilon} - c\hat{\vec{\alpha}}\cdot\hat{\vec{p}})\psi = -\hat{\beta}\ (m + m_{ad})c^2\psi \tag{5.6}$$

We can write the right part of the equation (5.7) through currents (3.13). Using the expression of mass term through the currents of electromagnetic representation and taking into account that $\omega = \dfrac{mc^2}{\hbar}$ we will obtain:

$$mc^2 \vec{E} = -i4\pi\hbar \vec{j}^e, \quad mc^2 \vec{H} = -i4\pi\hbar \vec{j}^m, \tag{5.7}$$

or, taking into account (4.2)

$$mc^2 \psi = -i4\pi\hbar \vec{j} \tag{5.7'}$$

where the matrix $j$, which in this case has the components $\{j_x^e, j_z^m, j_x^e, j_z^m\}$, we can name the current matrix.

Substituting (5.7) in (5.6'), we will obtain:

$$\left(\hat{\alpha}_o \hat{\varepsilon} - c\hat{\vec{\alpha}} \cdot \hat{\vec{p}}\right)\psi = -i\hat{\beta}\,4\pi\hbar(j_e + j_{ad}) \tag{5.8}$$

Comparing the above formulas we can make the following conclusions:
In the framework of CWED the external field acts as an addition to of the electron's own current, i.e. acts as some external current (we can assume that in the quantum form of the Dirac equation the external field can be considered as the addition to the mass of the electron);

The external field can be considered as a medium that has some polarization properties, which, obviously, can be characterized by the variable electric and magnetic permeability. In this case it is possible to express the additional current through the polarization vectors of medium as it is done in classical electrodynamics.

From the above following also that in framework of CWED the interaction of elementary particles can be considered as dispersion at their propagation in a medium, which consists from other particles (wave optics representation).

## 6.0. Connection of the de Broglie waves refraction index with Hamiltonian

It is not difficult to see, that in frameworks of CWED the equations (5.1) of interaction of the electron with other charged particles (or, in other words, the equations of the electron motion in the field of other particle) can be presented in form of the equations of the classical electrodynamics of medium:

$$\frac{1}{c}\frac{\partial \vec{E}}{\partial t} - rot\vec{H} = -\frac{4\pi}{c}\left(\vec{j}^e + \vec{j}_{ex}^e\right), \tag{6.1}$$

$$\frac{1}{c}\frac{\partial \vec{H}}{\partial t} + rot\vec{E} = \frac{4\pi}{c}\left(\vec{j}^m + \vec{j}_{ex}^m\right), \tag{6.2}$$

where $\vec{j}^e, \vec{j}^m$ are the electric and magnetic current densities of the particle, $\vec{j}_{ex}^e, \vec{j}_{ex}^m$ are the external current densities, which caused by the interaction of the given particle with other particles. In case if other particles form a medium as for example the physical vacuum, it can be presented using the electromagnetic theory of polarised medium [10,11,13,14]. In this case the external currents can be represented in the following way:

$$\vec{j}_{ex}^e = i\varepsilon_{ex}\vec{E}, \tag{6.3}$$

$$\vec{j}_{ex}^m = i\mu_{ex}\vec{H}, \tag{6.4}$$

where $\varepsilon_{ex}$ and $\mu_{ex}$ are permittivity and permeability of the external medium, i.e. of the external particles.

The Hamiltonian of Dirac's lepton theory (5.1) is following:

$$\hat{H} = c\hat{\vec{\alpha}} \cdot \hat{\vec{p}}\psi - \left[\hat{\beta} \, mc^2 + \left(\hat{\alpha}_o \varepsilon_{ex} - c\hat{\vec{\alpha}} \cdot \vec{p}_{ex}\right)\right]\psi, \qquad (6.5)$$

Using (4.2) we can obtain the CWED representation of (6.5), which we will conditionally write in the form:

$$\overline{H} \Leftrightarrow \pm rot(\vec{E},\vec{H}) \mp \frac{4\pi}{c}\left(\vec{j}^{e,m} + \vec{j}_{ex}^{e,m}\right), \qquad (6.5')$$

The expression (6.5') show that the connection of Hamiltonian with above currents (6.3) and (6.4) and correspondingly with the features of external medium $\varepsilon_{ex}$ and $\mu_{ex}$ exists.

Due to above result the Schroedinger equation with an external field can be written down through a "quantum" refraction index of medium. Conformity between electrodynamics of optical waves and electrodynamics of de Broglie waves is the most evident look for the stationary Schroedinger equation. Actually, the stationary Schroedinger equation:

$$\nabla^2 \psi + \frac{2m}{\hbar^2}(\varepsilon - \varepsilon_{int})\psi = 0, \qquad (6.6)$$

(where the energy $\varepsilon$ are Hamiltonian eigenvalues, $\varepsilon_{int} = e\varphi(r)$ is an interaction energy) is similar [17] to the optical wave equation, which determinates the light propagation in the medium whose refraction index changes in space from point to point:

$$\nabla^2 \psi + \left(\frac{2\pi \, n}{\lambda_0}\right)^2 \psi = 0, \qquad (6.7)$$

where $n = n(r) = \sqrt{\varepsilon_{ex}\mu_{ex}}$ is a refraction index, $\lambda_0$ is the length of wave in vacuum; and the optical wave length $\lambda = \frac{\lambda_0}{n}$ corresponds to the length of the de Broglie wave $\lambda = \frac{h}{p} = \frac{h}{\sqrt{2m(\varepsilon - \varepsilon_{int})}}$.

Since the elementary particles of CWED are the twirled electromagnetic waves [1,2], from the above follows that at their passage through the medium the refraction, diffraction, interference, and also absorption and division of these waves takes place, as for usual light waves (i.e. for linear photons). In this case the interconversion of particles at the collision can be considered as a dispersion of the curvilinear waves. Therefore, we can suppose that the dispersion matrix of the field theory performs the same role, as a dispersion matrix in optics.

## 7.0. Interaction Lagrangian of the non-linear Dirac-like equation of CWED

According to the energy-momentum conservation law, in the linear form we have:

$$\hat{\beta} \, mc^2 = \pm\left(\hat{\alpha}_0 \varepsilon_p - c\hat{\vec{\alpha}} \cdot \vec{p}_p\right), \qquad (7.1)$$

where in electromagnetic form we have:

$$\varepsilon_p = \int_0^\tau U \, d\tau = \frac{1}{8\pi}\int_0^\tau \left(\vec{E}^2 + \vec{H}^2\right) d\tau, \qquad (7.2)$$

$$\vec{p}_p = \int_0^\tau \vec{g} \, d\tau = \frac{1}{c^2}\int_0^\tau \vec{S} \, d\tau = \frac{1}{4\pi}\int_0^\tau [\vec{E}\times\vec{H}] \, d\tau, \qquad (7.3)$$

where in the general case the upper limit $\tau$ is equal to infinity. The same in quantum form we have:

$$\varepsilon_p = \frac{1}{8\pi}\int_0^\tau \psi^+\hat{\alpha}_0\psi \, d\tau, \quad \vec{p}_p = \frac{1}{4\pi}\int_0^\tau \psi^+\hat{\vec{\alpha}}\psi \, d\tau, \qquad (7.4)$$

Using the quantum form of $U$ and $\vec{S}$:

$$U = \frac{1}{8\pi}\psi^+\hat{\alpha}_0\psi, \qquad (7.5)$$

$$\vec{S} = -\frac{c}{8\pi}\psi^+\hat{\vec{\alpha}}\psi = c^2\vec{g}, \qquad (7.6)$$

and taking in to account that the free electron Dirac equation solution is the plane wave:

$$\psi = \psi_0 \exp[i(\omega t - ky)], \qquad (7.7)$$

we can write (7.5) and (7.6) in the next approximate form:

$$\varepsilon_p = U\,\Delta\tau = \frac{\Delta\tau}{8\pi}\psi^+\hat{\alpha}_0\psi, \qquad (7.8)$$

$$\vec{p}_p = \vec{g}\,\Delta\tau = \frac{1}{c^2}\vec{S}\,\Delta\tau = -\frac{\Delta\tau}{8\pi c}\psi^+\hat{\vec{\alpha}}\psi, \qquad (7.9)$$

The Lagrangian density of nonlinear equation is not difficult to obtain from the Lagrangian density of the linear Dirac equation [2]. By substituting (7.1) in (4.11) we obtain:

$$L_N = \psi^+\left(\hat{\varepsilon} - c\hat{\vec{\alpha}}\cdot\hat{\vec{p}}\right)\psi + \psi^+\left(\varepsilon_p - c\hat{\vec{\alpha}}\cdot\vec{p}_p\right)\psi, \qquad (7.10)$$

We suppose that the expression (7.10) represents the common form of the Lagrangian density of the nonlinear twirled electromagnetic wave equation.

Using (7.8) and (7.9) we can represent (4.11) in the approximate quantum form:

$$L_N = i\hbar\left[\frac{\partial}{\partial t}\left[\frac{1}{2}(\psi^+\psi)\right] - c\,\mathrm{div}(\psi^+\hat{\vec{\alpha}}\psi)\right] + \frac{\Delta\tau}{8\pi}\left[(\psi^+\psi)^2 - (\psi^+\hat{\vec{\alpha}}\psi)^2\right], \qquad (7.11)$$

By the normalizing $\psi$-function by the expression $L'_N = \frac{1}{8\pi\,mc^2}L_N$ and transforming (7.10) in the electrodynamics form, using equations (7.2) and (7.3), we will obtain from (7.10) the following approximate electromagnetic form:

$$L'_N = i\frac{\hbar}{2mc^2}\left(\frac{\partial U}{\partial t} + \mathrm{div}\,\vec{S}\right) + \frac{\Delta\tau}{mc^2}(U^2 - c^2\vec{g}^2), \qquad (7.12)$$

(note that correctly we must distinguish the complex-conjugate vectors $\vec{E}^*$ and $\vec{E}$, $\vec{H}^*$ and $\vec{H}$). It is not difficult to transform the second summand, using the known electrodynamics transformation:

$$(8\pi)^2(U^2 - c^2\vec{g}^2) = (\vec{E}^2 + \vec{H}^2)^2 - 4(\vec{E}\times\vec{H})^2 = (\vec{E}^2 - \vec{H}^2)^2 + 4(\vec{E}\cdot\vec{H})^2, \qquad (7.13)$$

Thus, taking into account that $L_D = 0$ and using (7.12) and (7.13), we obtain from (7.11) the following expression:

$$L'_N = \frac{1}{8\pi}\left(\vec{E}^2 - \vec{H}^2\right) + \frac{\Delta\tau}{(8\pi)^2 mc^2}\left[\left(\vec{E}^2 - \vec{H}^2\right)^2 + 4\left(\vec{E}\cdot\vec{H}\right)^2\right], \tag{7.14}$$

As we see, the approximate form of the Lagrangian density of the nonlinear equation of the twirled electromagnetic wave contains only the invariants of the Maxwell theory and is similar to the known Lagrangian density of the photon-photon interaction [18,19].

Let's now analyze the quantum form of the Lagrangian density (7.14). The equation (7.11) can be written in the form:

$$L_Q = \psi^+ \hat{\alpha}_\mu \partial_\mu \psi + \frac{\Delta\tau}{8\pi}\left[\left(\psi^+ \hat{\alpha}_0 \psi\right)^2 - \left(\psi^+ \hat{\vec{\alpha}}\, \psi\right)^2\right], \tag{7.15}$$

It is not difficult to see that the electrodynamics correlation (7.13) in quantum form has the known form of the Fierz correlation [20]:

$$\left(\psi^+ \hat{\alpha}_0 \psi\right)^2 - \left(\psi^+ \hat{\vec{\alpha}}\, \psi\right)^2 = \left(\psi^+ \hat{\alpha}_4 \psi\right)^2 + \left(\psi^+ \hat{\alpha}_5 \psi\right)^2, \tag{7.16}$$

Using (5.18) from (5.17) we obtain:

$$L_Q = \psi^+ \hat{\alpha}_\mu \partial_\mu \psi + \frac{\Delta\tau}{8\pi}\left[\left(\psi^+ \hat{\alpha}_4 \psi\right)^2 - \left(\psi^+ \hat{\alpha}_5 \psi\right)^2\right], \tag{7.17}$$

The Lagrangian density (7.17) coincides with the Nambu and Jona-Lasinio Lagrangian density [21], which is the Lagrangian density of the relativistic superconductivity theory. As it is known this Lagrangian density is used for the solution of the problem of the elementary particles mass appearance by the mechanism of the vacuum symmetry spontaneous breakdown [8].

As it is known [8, 22,23] the spontaneous breaking of symmetry consists in the fact that from a massless vector field, which has two spin states (such as the linear photon), and a massless scalar field $\phi$, a massive vector particle with three isospin projections appears. It is remarkable, that this break arises at the non-linear interaction of field $\phi$ with itself, whose self-action energy can be written down as potential

$$V(\phi) = \lambda^2 \left(|\phi|^2 - \eta^2\right)^2, \tag{7.18}$$

where $|\phi|^2$ is isoscalar, $\lambda$ is a dimensionless parameter, $\eta$ is a parameter with the dimension of the mass. It is easy to see the analogies of the spontaneous breaking of symmetry with description of the pair production process in CWED [2].

## 8.0. The general case of non-linear field theory Lagrangian and Hamiltonian of interaction

As we noted, the Hamiltonian is the full system energy. For example, in case of interaction of two elementary particles this energy consists of the sum of three parts: the researched particle's own energy, the energy of its interaction with another particle and the energy of the interaction of a researched particle with physical vacuum (the last term corresponds to the account of the polarization of physical vacuum).

As we have shown in general case the CWED is the non-linear theory. Therefore the Lagrangian of a system must contain all possible terms with the invariants of electromagnetic field.

In this case the Lagrangian of the non-linear field theory can be written as some function of the field invariants:

$$\overline{L} = f_L(I_1, I_2), \qquad (8.1)$$

where $I_1 = (\vec{E}^2 - \vec{H}^2)$, $I_2 = (\vec{E} \cdot \vec{H})$ are the invariants of electromagnetic field theory.

Hamiltonian is fully defined through the Lagrangian of a system. If the function (8.1) is known, using the formula (1.13), it is easy to calculate the Hamiltonian of the system, which will be now, obviously, the function of the various powers of electromagnetic field vectors

$$\overline{H} = f_H(\vec{E}, \vec{H}), \qquad (8.2)$$

Apparently, for each problem the function $f_L$ has its special form, which is unknown (the same happens with the form of the function $f_H$).

It has appeared, that the approximate form of the function $f_H$ can be found on the basis of Schroedinger (or Dirac) wave equation, using the so-called perturbation method.

Generally this procedure has the following form. It is supposed, that there is an expansion of the function $f_H$ in Taylor – MacLaurent power series with unknown expansion coefficient. Then the problem is reduced to the calculation of these coefficients. The solution is searched for each term of expansion separately, starting with the first. Usually for the first term of expansion this is a problem for a free particle, whose solution is already known. Then using the equation with the two first terms, we find the coefficient of the second term of expansion. Then, using the equation for the three first terms, we find the coefficient for the third term of expansion, etc. As it is known by this method it is possible to obtain the solution with any desirable accuracy.

As it is known in case of functions of two variables $\xi = f(x, y)$ the Taylor – MacLaurent power series nearly to a point $(x_0, y_0)$ is:

$$f(x, y) = f(x_0, y_0) + \sum_{k=1}^{n} \frac{1}{k!}\left((x-x_0)\frac{\partial}{\partial x} + (y-y_0)\frac{\partial}{\partial y}\right)^k f(x_0, y_0) + O(\rho^n), \qquad (8.3)$$

where $\rho = \sqrt{(x-x_0)^2 + (y-y_0)^2}$,

$$\left((x-x_0)\frac{\partial}{\partial x} + (y-y_0)\frac{\partial}{\partial y}\right) f(x_0, y_0) \equiv (x-x_0)\frac{\partial f(x_0, y_0)}{\partial x} + (y-y_0)\frac{\partial f(x_0, y_0)}{\partial y}, \qquad (8.4)$$

$$\left((x-x_0)\frac{\partial}{\partial x} + (y-y_0)\frac{\partial}{\partial y}\right)^2 f(x_0, y_0) \equiv (x-x_0)^2 \frac{\partial f^2(x_0, y_0)}{\partial x^2} + \\ + 2(x-x_0)(y-y_0)\frac{\partial^2 f(x_0, y_0)}{\partial x \partial y} + (y-y_0)^2 \frac{\partial f^2(x_0, y_0)}{\partial y^2}, \qquad (8.5)$$

Etc.

In a case when $x_0 = 0$, $y_0 = 0$ we obtain the MacLaurent series.

Obviously, for the most types of the functions $f_L(I_1, I_2)$ the expansion contains approximately the same set of the terms, which distinguish only by the constant coefficients, any of which can be equal to zero (as an example of the expansion it is possible to point out the expansion of the quantum electrodynamics Lagrangian for particle at the present of physical vacuum [18, 19, 24]). Generally the expansion looks like:

$$L_M = \frac{1}{8\pi}(\vec{E}^2 - \vec{B}^2) + L', \qquad (8.6)$$

22where

$$L' = \alpha \left(\vec{E}^2 - \vec{B}^2\right)^2 + \beta \left(\vec{E} \cdot \vec{B}\right)^2 + \gamma \left(\vec{E}^2 - \vec{B}^2\right)\left(\vec{E} \cdot \vec{B}\right) + \xi \left(\vec{E}^2 - \vec{B}^2\right)^3 + \zeta \left(\vec{E}^2 - \vec{B}^2\right)\left(\vec{E} \cdot \vec{B}\right)^2 + \ldots, \quad (8.7)$$

is the part which is responsible for the non-linear interaction (here $\alpha, \beta, \gamma, \xi, \zeta, \ldots$ are constants)

Corresponding Hamiltonian will be defined as follows:

$$\overline{H} = \sum_i E_i \frac{\partial L}{\partial E_i} - L = \frac{1}{8\pi}\left(\vec{E}^2 + \vec{B}^2\right) + \overline{H}', \quad (8.8)$$

where the Hamiltonian part responsible for non-linear interaction is:

$$\overline{H}' = \alpha \left(\vec{E}^2 - \vec{B}^2\right)\left(3\vec{E}^2 - \vec{B}^2\right) + \beta \left(\vec{E} \cdot \vec{B}\right)^2 + \xi \left(\vec{E}^2 - \vec{B}^2\right)\left(5\vec{E}^2 + \vec{B}^2\right) + \zeta \left(3\vec{E}^2 - \vec{B}^2\right)\left(\vec{E} \cdot \vec{B}\right)^2 + \ldots, \quad (8.9)$$

It is not difficult to obtain the quantum representation of non-linear theory Hamiltonian (8.9). Replacing the electromagnetic wave field vectors by quantum wave function according to (4.2), we will obtain a series of type

$$\hat{\overline{H}} = \left(\psi^+ \hat{\alpha}_0 \psi\right) + \sum c_{1i}\left(\psi^+ \hat{\alpha}_i \psi\right)\left(\psi^+ \hat{\alpha}_j \psi\right) + \sum c_{2i}\left(\psi^+ \hat{\alpha}_i \psi\right)\left(\psi^+ \hat{\alpha}_j \psi\right)\left(\psi^+ \hat{\alpha}_k \psi\right) + \ldots, \quad (8.10)$$

where $\hat{\alpha}_i, \hat{\alpha}_j, \hat{\alpha}_k$ are Dirac matrixes, $c_i$ are coefficients of expansion.

As we see, the terms of Lagrangian and Hamiltonian series contain enough limited number of the same elements, such $\left(\vec{E}^2 + \vec{B}^2\right)$, $\left(\vec{E} \cdot \vec{B}\right)^2$, $\left(\vec{E}^2 - \vec{B}^2\right)$ as well as some other. It is possible to assume, that each element of series has some constant physical sense. In this case it is possible to see the analogy with expansion of fields on the electromagnetic moments (2.23), or with decomposition of a S-matrix on the elements [18], each of which accords to the particularities of interaction of separate particles.

## Conclusion

The above results show that in framework of CWED there is the uniform description of fields' interaction. As we can see the descriptions of the interactions, used in different sections of physics (mechanics, classical electrodynamics, quantum electrodynamics, etc.), are equivalent. It appears also that the interaction can be expressed by field strengths, potentials, wave functions, currents, energy-momentum tensor, stress, medium polarization index etc. This allows us to speak about CWED as about the unified field theory.